\documentstyle[emulateapj,psfig]{article}
\makeatletter
\newenvironment{inlinefigure}{%
\def\@captype{figure}%
\noindent\begin{minipage}{0.999\linewidth}\begin{center}}
{\end{center}\end{minipage}\smallskip}
\makeatother

\lefthead{Cowie et al.}
\begin{document}
\title{
$2-8$~keV X-ray number counts determined from Chandra blank field
observations
}
\author{L.\,L.\ Cowie,$\!$\altaffilmark{1}
G.\,P.\ Garmire,$\!$\altaffilmark{2}
M.\,W.\ Bautz,$\!$\altaffilmark{3}
A.\,J.\ Barger,$\!$\altaffilmark{4,5,1}
W.\,N.\ Brandt,$\!$\altaffilmark{2}
A.\,E.\ Hornschemeier$\!$\altaffilmark{2}
}

\altaffiltext{1}{Institute for Astronomy, University of Hawaii,
2680 Woodlawn Drive, Honolulu, HI 96822}
\altaffiltext{2}{Department of Astronomy \& Astrophysics,
525 Davey Laboratory, The Pennsylvania State University,
University Park, PA 16802}
\altaffiltext{3}{Center for Space Research, Massachusetts Institute
of Technology, Cambridge, MA 02139}
\altaffiltext{4}{Department of Astronomy, University of Wisconsin-Madison,
475 North Charter Street, Madison, WI 53706}
\altaffiltext{5}{Department of Physics and Astronomy,
University of Hawaii, 2505 Correa Road, Honolulu, HI 96822}

\slugcomment{The Astrophysical Journal Letters in press.}

\begin{abstract}

We present the $2-8$~keV number counts
from the 1~Ms {\it Chandra} observation of the
{\it Chandra} Deep Field North (CDF-N). We combine these
with the number counts 
from a 78~ks exposure
of the Hawaii Survey Field SSA22 and with the number counts obtained
in independent analyses of the CDF-S and the Hawaii Survey Field
SSA13 to determine the number counts from $2\times 10^{-16}$
to $10^{-13}$~erg~cm$^{-2}$~s$^{-1}$. Over this flux
range the contribution to the X-ray background
is $1.1\times 10^{-11}$~erg~cm$^{-2}$~s$^{-1}$~deg$^{-2}$.
When the contributions above $10^{-13}$~erg~cm$^{-2}$~s$^{-1}$
from {\it BeppoSAX} or {\it ASCA} observations are included, the
total rises to $1.4\times 10^{-11}$~erg~cm$^{-2}$~s$^{-1}$~deg$^{-2}$.
However, there appears to be substantial field-to-field
variation in the counts in excess of the statistical
uncertainties. When the statistical and flux calibration
uncertainties (in both the background and source measurements) 
are taken into account, as much as 
$0.5\times 10^{-11}$~erg~cm$^{-2}$~s$^{-1}$~deg$^{-2}$ 
could still be present in an unresolved component.
\end{abstract}


\section{Introduction} \label{intro}

The X-ray background (XRB) was the first cosmic background  
detected (\markcite{giacconi62}Giacconi et al.\ 1962) and
has been extensively characterized
(\markcite{fabian92}Fabian \& Barcons 1992).
Its photon intensity $P(E)$, where $E$ is the photon 
energy in keV and $P(E)$ has units
[photons~cm$^{-2}$~s$^{-1}$~keV$^{-1}$~sr$^{-1}$],
can be approximated in the $1-15$~keV range by 
$P(E)=AE^{-\Gamma}$ where $\Gamma\simeq 1.4$
(e.g., \markcite{marshall80}Marshall et al.\ 1980; 
\markcite{gendreau95}Gendreau et al.\ 1995;
\markcite{chen97}Chen, Fabian, \& Gendreau 1997). 
However, the XRB normalization is still somewhat uncertain
with values in the $2-8$~keV band lying between
$1.3\times 10^{-11}$
(\markcite{marshall80}Marshall et al.\ 1980) and
$1.8\times 10^{-11}$~erg~cm$^{-2}$~s$^{-1}$~deg$^{-2}$
(\markcite{vecchi99}Vecchi et al.\ 1999).

After the discovery of the XRB there was 
controversy over whether it arose from a superposition 
of discrete sources or from thermal bremsstrahlung from a 
hot intergalactic gas (e.g., \markcite{field72}Field\ 1972).
We now know that the XRB cannot 
originate in an uniform hot intergalactic medium since the absence of
a strong Compton 
distortion in the cosmic microwave background spectrum
puts a stringent upper 
limit ($\sim 10^{-4}$) on such a contribution
(\markcite{wright94}Wright et al. 1994).
However, there may be other sources of hard X-ray photons
(e.g., \markcite{abazajian01}Abazajian, Fuller, \& Tucker 2001)
which could contribute to the XRB
without producing a Compton distortion, so
constraining any residual diffuse background is
of considerable interest. 

The sources contributing the bulk of the $2-8$~keV XRB were not 
found prior to the launch of the {\it Chandra X-ray Observatory}.
While the soft ($0.5-2$~keV) background was largely resolved
into sources by the {\it ROSAT} satellite
(\markcite{has98}Hasinger et al.\ 1998), the spectra
were too steep to account for the flat XRB spectrum. 
With the launch of {\it Chandra} and 
the {\it XMM-Newton Observatory} the situation has evolved rapidly. 
Early $100-300$~ks {\it Chandra} observations
determined the number counts in the $2-8$~keV band
above $10^{-15}$~erg~cm$^{-2}$~s$^{-1}$
(\markcite{mushotzky00}Mushotzky et al.\ 2000;
\markcite{giacconi01}Giacconi et al.\ 2001;
\markcite{tozzi01}Tozzi et al.\ 2001) and resolved
the majority of the $2-8$~keV XRB.
\markcite{has01}Hasinger et al. (2001)
have reported parallel results with {\it XMM-Newton}.
Now the two 1~Ms {\it Chandra} exposures of the
CDF-N (\markcite{brandt01}Brandt et al.\ 2001, hereafter B01)
and the {\it Chandra} Deep Field South (CDF-S;
\markcite{giacconi02}Giacconi et al.\ 2002, hereafter G02; 
\markcite{campana01}Campana et al.\ 2001, hereafter C01) 
have extended the counts to 
$2\times 10^{-16}$~erg~cm$^{-2}$~s$^{-1}$.
However, there is considerable spread (about 40\%)
in the normalizations, with the counts in the SSA13
(\markcite{mushotzky00}Mushotzky et al.\ 2000) and
CDF-N (B01) fields being higher than the CDF-S counts
(G02; C01) and the {\it XMM-Newton}
Lockman Hole counts (\markcite{has01}Hasinger et al.\ 2001).

Here we provide a more detailed analysis of
the number counts in the CDF-N,
modelling the effects of Eddington bias and
incompleteness.  We also analyze a new 78~ks observation of
the SSA22 field. We combine our counts with those
measured by \markcite{mushotzky00}Mushotzky et al.\ (2000) from the
100~ks observation of the SSA13 field and those
measured by C01 from the 1~Ms observation of the CDF-S to 
average over a spread of fields. We are currently analyzing
the latter two fields using the same methods described here
(L.L. Cowie et al., in preparation),
but we note that similar number counts have been obtained
for the CDF-S using very different methods (G02; C01).

We break with the tradition of X-ray number counts by 
using differential and not cumulative counts. 
Differential counts have many advantages
in the statistical 
independence of the data points and in the ease with which breaks
and shape changes may be seen (c.f., \markcite{jauncey75}Jauncey 1975). 

\section{X-ray Data}


The observations of the CDF-N are presented in B01,
where the $2-8$~keV image and the exposure map are described. 
We follow B01 and use $2-8$~keV fluxes since there is 
little sensitivity above 8~keV in the data.
When we use the data of other authors, we convert  
to this band using consistent assumptions for 
the spectral index. 

An important issue is the choice of spectrum for
converting counts to flux. 
B01 converted using the $\Gamma$ value determined 
from the ratio of the hard to soft band counts 
for each individual source. Others have 
converted using a single $\Gamma$ chosen to match 
the expected average spectrum of the sources. G02 and C01 used
$\Gamma = 1.4$, corresponding to the shape of the XRB,
while \markcite{mushotzky00}Mushotzky et al.\ (2000) and 
\markcite{barger01}Barger et al.\ (2001) used $\Gamma = 1.2$, 
based on the argument that the average spectra at these 
fluxes must be slightly harder than that of the XRB once
allowance is made for the contribution of brighter sources
(which, on average, are softer than the XRB). None of these
methods is fully satisfactory for individual sources
whose spectra may be more complex than a simple power-law.
We adopt the $\Gamma = 1.2$ method
which should give the best {\it average} conversion and
hence the best estimate of the contribution of the
sources to the XRB. This 
$\Gamma$, together with the 
$N_H=1.6\times 10^{20}$~cm$^{-2}$ of Galactic absorption, gives 
a conversion of counts to flux of 
$2.19\times 10^{-11}$~erg~cm$^{-2}$
for the charge transfer inefficiency corrected images of B01.
The $\Gamma = 1.4$ conversion would reduce the fluxes
by 4\%, while use of individual $\Gamma$s would increase them 
by 13\%. In combining the new counts with published 
number counts, we have applied the appropriate 
conversion.

There are various techniques for source
selection, including sliding cells, wavelet based
algorithms (B01; C01), and the optimal 
matched filter method (\markcite{vikhlinin95}Vikhlinin et al.\ 1995). 
In the present analysis we use a matched aperture, which
consists of a circular aperture of specified diameter
and a background subtraction region which is
an annulus $2-3$ times the aperture radius.
We determined the background from a cleaned image
in which all known sources were masked out with
artificial noise appropriate to the local exposure time. 
The background-subtracted source counts were determined by
centering apertures on each pixel in the image 
and finding the local maxima. In each
case if there was a secondary maximum within $2''$
of the primary, we combined the two sources.
We determined
the average background per aperture and
selected only sources whose background-subtracted
counts lay above the level where Poisson noise in the average
background within the apertures would provide
a probability of one false detection in the area. 

We first analyzed the central 
area of the CDF-N image lying within $2.5'$ of the optical
axis. Here we used a $2''$ aperture
selected to optimize the recovery (based on the Monte
Carlo simulations described below). 
This  gave an average number of background counts of
$1.8$ per aperture; thus, we selected sources with background-subtracted
counts lying above $9.0$.
We next analyzed the  $6.5'$ radius area around the optical
axis (including the previously analyzed $2.5'$ area) using  a $4''$
diameter aperture and selected sources with background-subtracted
counts above $14.7$. 
The regions have fairly uniform exposure times, 
which justifies the use of the average background. However, 
to test this, and also to check that the interchip regions, where 
the properties of the sources might be unusual, were not
affecting the counts, we reran the search, clipping out 
the regions with lower exposure times. This had no significant effect 
on the number counts.
%
%
\begin{inlinefigure}
\psfig{figure=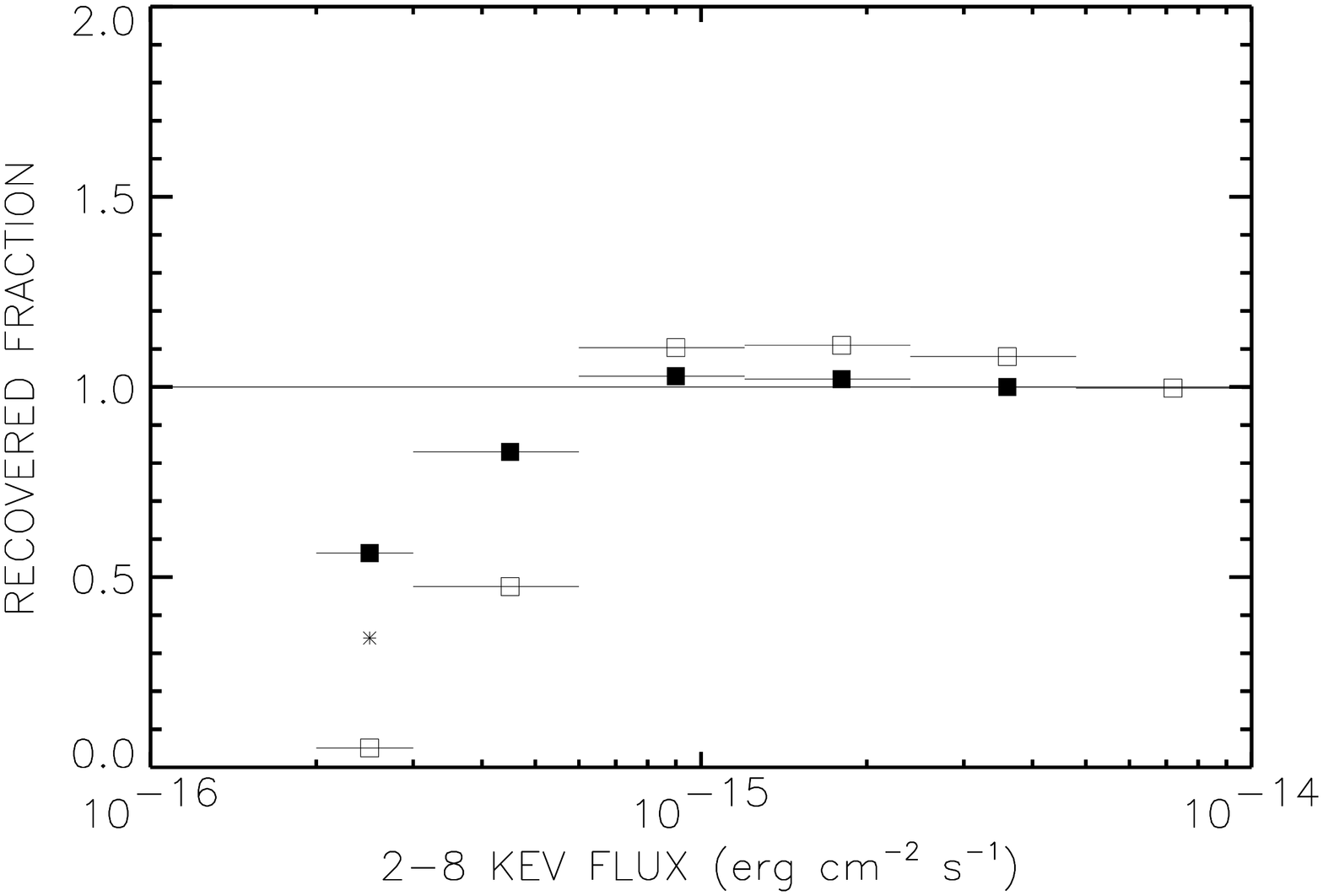,angle=90,width=3.5in}
\vspace{6pt}
\figurenum{1}
\caption{The ratio of the measured number counts
to the input number counts in the $2.5'$ radius
(solid boxes) and $6.5'$ radius (open boxes)
regions. The width of each bin is shown by the short horizontal
lines. Vertical error bars are smaller than the filled symbols.
At high fluxes the retrieved counts precisely match the input
counts. At intermediate fluxes Eddington
bias becomes important and the retrieved number of sources exceeds
the input counts. In the lowest flux bins, source recovery
becomes incomplete. The corrected area is largest
in the $6.5'$ region analysis except in the lowest bin
where the star shows the recovered fraction in the $6.5'$
region times the relative area of the two regions.
}
\label{fig1}
\addtolength{\baselineskip}{10pt}
\end{inlinefigure}

We used the source counts determined in the $4''$ diameter aperture,
with the subtracted background determined in the $8''-12''$ annulus, 
to get the fluxes of our sample. These counts were converted
to count rates using the B01 exposure map.
We determined the enclosed energy correction 
by measuring the ratio of the counts in a $20''$
diameter aperture relative to the $4''$ diameter aperture for 
all sources with fluxes above $10^{-15}$~erg~cm$^{-2}$~s$^{-1}$
and fitting this with a third order polynomial as a function
of the off-axis radius.  There are no extended sources
included in this computation. The enclosed energy correction
agrees well with the standard ACIS PSF correction.

The flux calibration was tested by comparing the detected sources with 
the B01 catalog converted to $\Gamma=1.2$. Above a flux
limit of $3.5\times 10^{-16}$~erg~cm$^{-2}$~s$^{-1}$, the sample
within the $6.5'$ radius region contains 152 sources and B01 141, 
of which 134 are common.
For the sources in the B01 catalog above $5\times
10^{-16}$~erg~cm$^{-2}$~s$^{-1}$, the mean fluxes agree to 1\% and the
dispersion of the fluxes, when multiple sources are excluded, is 8\%.  Number
counts generated from the two catalogs are nearly indistinguishable above this
flux. At fainter fluxes the catalogs become more disjoint, reflecting
the different source selection procedures, but the fluxes of
the overlapping sample continue to agree.

The effects of incompleteness and bias were measured using 
Monte Carlo simulations in which a number of model X-ray images were
created with noise properties matching that of the 1 Ms 
field and with sources drawn from parameterized descriptions
of the number counts. 100 such images
were generated and analyzed in the same way as the actual
X-ray image of the 1~Ms field.  The artificial images were 
generated by masking all known sources from the real 1~Ms image
and substituting artificial noise.
Sources were then randomly drawn from the number
counts model and added to this artificial image. To 
match the point spread function (PSF) the template 
sources were drawn from the brightest source in the real
1~Ms image in a $1'$ width 
annulus centered on the input position. This template was 
background-subtracted and flux-scaled to produce the correct 
input flux. The  
source identification and cataloging procedures were now run on
the artificial image to generate 
the output counts. The output counts averaged over
the 100 simulations were then compared with 
the input counts to determine the recovered fraction, which 
could then be used to correct the measured number counts. The 
input number counts model was next adjusted to match the 
observed number counts and the process repeated. The 
recovered fractions are shown in 
Figure~\ref{fig1} and illustrate the effect of Eddington
bias (the recovered sources are brightened, which causes the observed 
counts to be high at intermediate fluxes) and the effect
of incompleteness in the recovery at the lowest fluxes.

For fluxes between $2\times 10^{-16}$
and $3\times 10^{-16}$~erg~cm$^{-2}$~s$^{-1}$,
we used the $2.5'$ radius sample and its corrected area.
Between $3\times 10^{-16}$
and $5\times 10^{-15}$~erg~cm$^{-2}$~s$^{-1}$,
we used the $6.5'$ sample and its corrected area.
Above $5\times 10^{-15}$~erg~cm$^{-2}$~s$^{-1}$,
we used the entire CDF-N area lying within $10'$
of the optical axis, an area of 294 square arcminutes.
At these large off-axis angles, simulations are
not straightforward because of the complex
PSFs which arise from the multiple positions and
roll angles at which the data were taken. Here the fluxes were 
taken from the B01 catalog (converted to $\Gamma = 1.2$),
which carefully calculated the total counts for the bright 
objects. The incompleteness and bias corrections were determined 
by comparing the raw number counts with those in the $6.5'$ radius
region. $5\times 10^{-15}$~erg~cm$^{-2}$~s$^{-1}$
lies well above the flux at which
such effects become significant.


The 78~ks SSA22 field was analyzed using the standard reduction tools. 
A complete list of sources lying in the S2 and S3 chips and their
fluxes were generated with WAVDETECT, using the assumptions of the present
paper. More details can be found in 
M. W. Bautz et al., in preparation. The total area used was
$0.035$ square degrees. Only sources lying above 
$6\times 10^{-15}$~erg~cm$^{-2}$~s$^{-1}$\ 
(above which incompleteness and bias corrections are small)
are included in the present analysis. 

\section{Number Counts} 
\label{numb}

We combined the fields by forming a total list of sources
and determining the combined incompleteness and bias-corrected areas 
sampled at each flux. For the CDF-S we included the entire
sample from C01, while for the SSA13
counts from \markcite{mushotzky00}Mushotzky et al.\ (2000)
we restricted ourselves to sources with
fluxes above $4\times 10^{-15}$~erg~cm$^{-2}$~s$^{-1}$
to avoid any issues of bias and incompleteness which
might be present at lower fluxes. The final
sample consists of 373 sources with fluxes between
$2\times 10^{-16}$ and 
$1.3\times 10^{-13}$~erg~cm$^{-2}$~s$^{-1}$.
The total corrected area ranges from 40 square
arcminutes at the faint flux end to 0.25 square degrees at 
bright fluxes.  The differential counts are the sum of the 
inverse areas of the sources in the flux interval ${\Delta S}$ 
divided by the width: 
\begin{equation}
n(S) = {{\sum (1/A_i)} \over {\Delta S}}
\end{equation}
We have normalized to a unit flux of 
$10^{-15}$~erg~cm$^{-2}$~s$^{-1}$.
In Fig.~\ref{fig2} we present the differential number counts
per square degree per unit flux (filled squares) with $1\sigma$
uncertainties from the Poisson error in the number of  
sources. We also show the number counts from only the
CDF-N and Hawaii fields (open triangles)
and the CDF-S counts from C01 (open diamonds).
There is a substantial difference between these two determinations
of the number counts at fluxes below $10^{-15}$~erg~cm$^{-2}$~s$^{-1}$. 
This difference could reflect the different methods of analysis
(indeed, it occurs at fluxes where modelling
corrections are required);  however, all the methods of
analysis so far applied to the fields have shown these same
differences (B01; \markcite{giacconi01b}Giacconi et al.\ 2001b; 
C01), and it appears more probable that they result from
field-to-field variations.
%
%
\begin{inlinefigure}
\psfig{figure=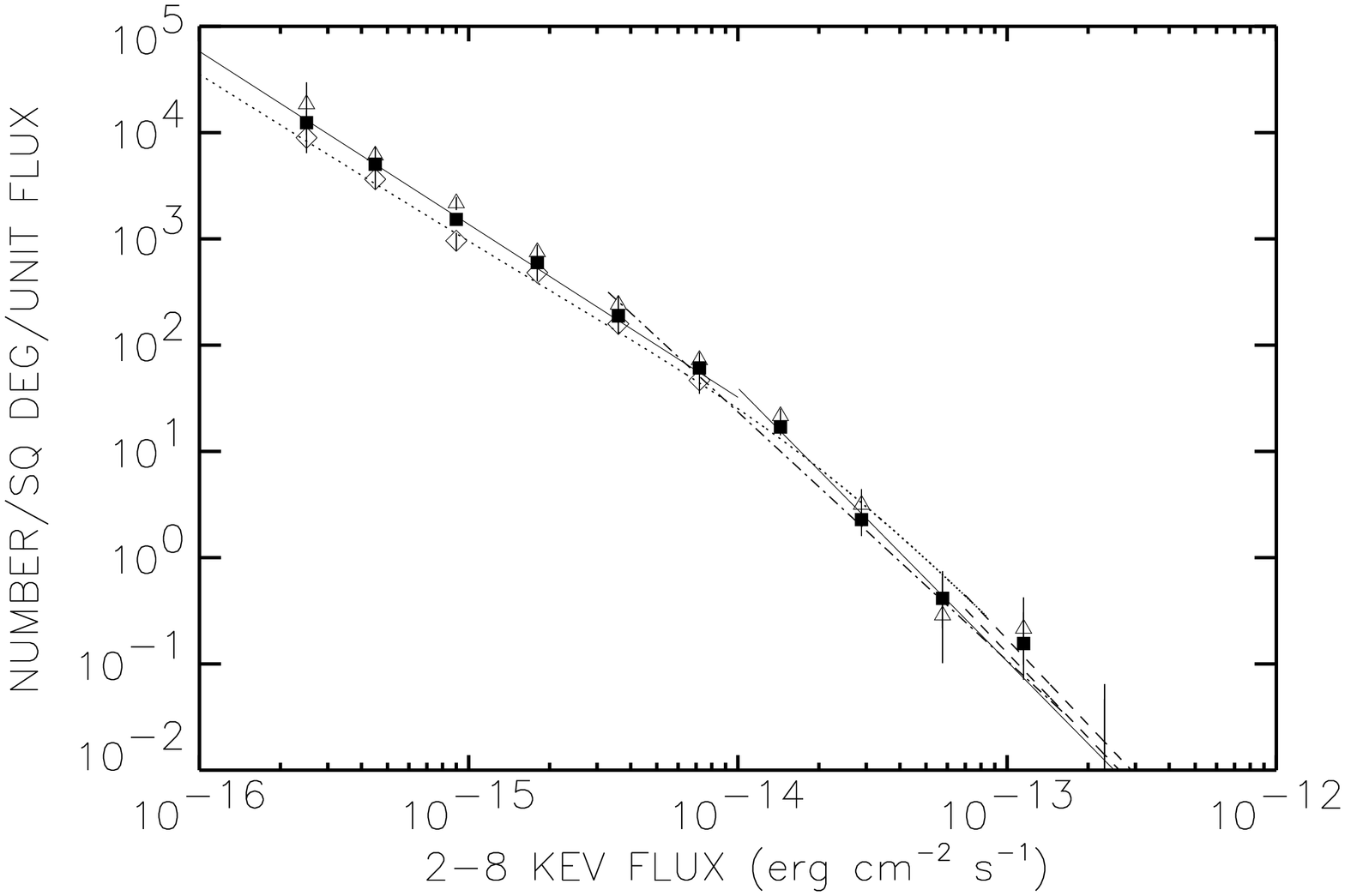,angle=90,width=3.5in}
\vspace{6pt}
\figurenum{2}
\caption{The combined differential counts from the four
fields (solid boxes) with $1\sigma$ uncertainties.
The y-axis units are relative to a unit flux of
$10^{-15}$~erg~cm$^{-2}$~s$^{-1}$.
The open diamonds show the C01 CDF-S counts, and the open triangles
show the combined CDF-N, SSA22, and SSA13 counts.
The solid lines show the power-law fits to
the composite counts over the flux ranges above and below
$10^{-14}$~erg~cm$^{-2}$~s$^{-1}$, while the dashed lines
show the {\it ASCA} counts of della Cecca et al.\ (2000) and
the {\it BeppoSAX} counts of Giommi et al.\ (2000).
The dot-dash line shows Baldi et al. (2001)'s fit to
their {\it XMM-Newton} counts.
The dotted line shows C01's fit to the CDF-S and
{\it ASCA} data.
}
\label{fig2}
\addtolength{\baselineskip}{10pt}
\end{inlinefigure}

The number counts cannot be fit by a single
power-law but are well fit by a broken power-law.
An area weighted maximum likelihood fit to the counts from
all four fields, utilizing the bias-corrected areas
(\markcite{murdoch73}Murdoch, Crawford, \& Jauncey 1973)
gives a fit above
$10^{-14}$~erg~cm$^{-2}$~s$^{-1}$ of
\begin{equation}
n(S)=(39\pm5) \times (S/10^{-14})^{-2.57\pm0.22}
\label{eq1}
\end{equation}

\noindent
and below $10^{-14}$~erg~cm$^{-2}$~s$^{-1}$ of
\begin{equation}
n(S)=(32\pm2) \times (S/10^{-14})^{-1.63\pm0.05}.
\label{eq2}
\end{equation}

\noindent
The normalizations are determined by matching the total 
number of objects in each range, and the uncertainties are $1\sigma$.  
The units of $n(S)$ are number per square degree per
$10^{-15}$~erg~cm$^{-2}$~s$^{-1}$.
The two power-laws intercept 
at a flux of $1.2\times 10^{-14}$~erg~cm$^{-2}$~s$^{-1}$.
A fit with the uncorrected areas gives a
slightly steeper slope at the faint end, consistent with 
the analytic expectation when uncorrected areas are used instead of
bias-corrected areas
(e.g., \markcite{schmitt86} Schmitt and Maccacaro 1986).
Fitting to only the CDF-N+Hawaii field counts gives a similar
faint end slope (${-1.61\pm0.06}$) but higher
normalization (${45\pm4}$), which, on integration, 
closely matches the cumulative $2-8$~keV counts of
B01.
C01, using only the CDF-S data and force fitting the bright
end of the flux counts to the {\it ASCA} slope, inferred
a faint flux slope of $-1.58\pm 0.03$ and a break point
of $1.7\times 10^{-14}$~erg~cm$^{-2}$s$^{-1}$
(the dotted line in Fig.~\ref{fig2}).
A systematic effect which could change the slope is the
progressive hardening of the sources as
the flux drops. 
The median ${\Gamma}$ in B01 drops from
$1.4$ at $10^{-14}$~erg~cm$^{-2}$~s$^{-1}$ to
$0.9$ at the faintest fluxes. 
Using these ${\Gamma}$~s to compute the flux conversion
increases the faint end slope
to $-1.68$. 

Cross calibration is a concern in comparing with counts
from other satellites.
Tests using a point
source common to {\it ASCA} and {\it Chandra}
give agreement at about the 10\% level for these
satellites (\markcite{barger01}Barger et al. (2001))
A more detailed analysis has recently been given
by \markcite{snow01}Snowden et al. (2001)
who found that the {\it XMM} and {\it Chandra} ACIS-S3 data
agree to about 5\% in the $2-10$~keV band and that the 
{\it ASCA} and {\it Chandra}
fluxes agree to about 10\%.
The bright counts presented here match extremely
well to the counts obtained by Baldi et al. (2001)
using {\it XMM-Newton} (dash-dot line in Fig.~\ref{fig2})
and lie only slightly below the {\it ASCA} 
counts of \markcite{ueda99}Ueda et al.\ (1999) and 
\markcite{della00}della Ceca et al.\ (2000) 
and the {\it BeppoSAX} counts 
of \markcite{giommi00}Giommi et al.\ (2000)
(dashed lines in Fig.~\ref{fig2}).

\section{Discussion}

Fig.~\ref{fig3} shows the contribution to the XRB versus flux.
The peak contribution arises near the break 
at $1.4 \times 10^{-14}$~erg~cm$^{-2}$~s$^{-1}$.
The total contribution is 
$1.1\times 10^{-11}$~erg~cm$^{-2}$~s$^{-1}$~deg$^{-2}$
over the flux range $2\times 10^{-16}$
to $10^{-13}$~erg~cm$^{-2}$~s$^{-1}$.
Extrapolating the power-law fit would add a further 
$0.2\times 10^{-11}$~erg~cm$^{-2}$~s$^{-1}$~deg$^{-2}$
above $10^{-13}$~erg~cm$^{-2}$~s$^{-1}$, similar to the
value obtained from the {\it BeppoSAX}
number counts. Use of the highest {\it ASCA} counts would 
only increase this to $0.3\times 10^{-11}$~erg~cm$^{-2}$~s$^{-1}$~deg$^{-2}$.
Extrapolating the faint end flux counts below 
$2\times 10^{-16}$~erg~cm$^{-2}$~s$^{-1}$
would add $0.2\times 10^{-11}$~erg~cm$^{-2}$~s$^{-1}$~deg$^{-2}$
to the total. For the {\it HEAO-1} normalization
the observed contribution from sources above
$2\times 10^{-16}$~erg~cm$^{-2}$~s$^{-1}$
exceeds the background, while for
the {\it BeppoSAX} and {\it ASCA} normalizations it is
approximately 80\% of the XRB. This would rise
to 90\% if we include the extrapolated faint end 
flux contribution.
 
The field-to-field differences can clearly be seen in Fig.~\ref{fig3}, 
with the CDF-S and SSA22 lying below the CDF-N and
SSA13. Between $2\times 10^{-16}$
and $10^{-14}$~erg~cm$^{-2}$~s$^{-1}$ the
CDF-N contributes $0.81\times 10^{-11}$~erg~cm$^{-2}$~s$^{-1}$~deg$^{-2}$,
while the CDF-S contributes $0.54\times 10^{-11}$.
This 40\% difference is substantially 
above the Poisson noise expected from the
number of sources. 
%
%
\begin{inlinefigure}
\psfig{figure=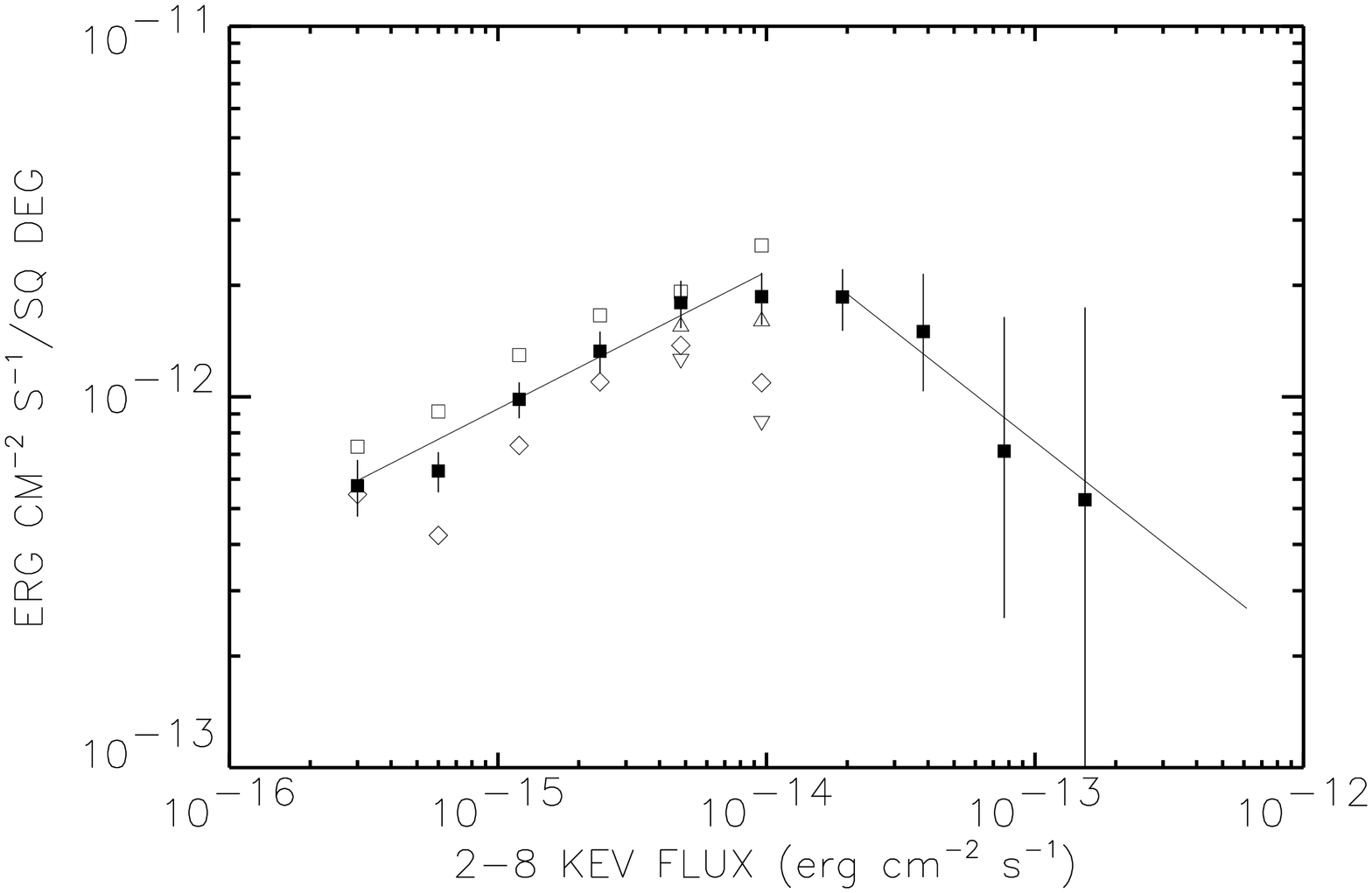,angle=90,width=3.4in}
\vspace{6pt}
\figurenum{3}
\caption{The contribution to the XRB versus
flux. The solid boxes are the
measured values in the combined
sample. The lines show the values
from the power-law
fits. The open boxes show the 
CDF-N, the open diamonds 
the CDF-S, the open upward pointing triangles
the SSA13 field, and the open downward pointing triangles
the SSA22 field. The individual fields are
shown only below $10^{-14}$~erg~cm$^{-2}$~s$^{-1}$
where the error bars are small.
}
\label{fig3}
\addtolength{\baselineskip}{10pt}
\end{inlinefigure}

Given these variations and the uncertainties 
in the flux calibration of both the counts and the background, 
we estimate the minimum contribution of the resolved sources to the XRB to be
$1.3\times 10^{-11}$~erg~cm$^{-2}$~s$^{-1}$~deg$^{-2}$.
Comparing this number to the maximum measured value of the XRB,
$1.8\times 10^{-11}$~erg~cm$^{-2}$~s$^{-1}$~deg$^{-2}$
(\markcite{vecchi99}Vecchi et al.\ 1999), we find a maximum residual of
$0.5\times 10^{-11}$~erg~cm$^{-2}$~s$^{-1}$~deg$^{-2}$.
The true value of any diffuse component is likely to be considerably lower.

\acknowledgements
We thank Sergio Campana for supplying the data on the CDF-S.
Support came from NASA grant DF1-2001X (LLC, PI), 
NSF grant AST-0084816 (LLC), NASA grant NAS 8-38252 (GPG, PI),
University of Wisconsin Research Committee funds
granted by the Wisconsin Alumni Research Foundation (AJB),
NSF grant AST-0084847 (AJB), NSF CAREER award AST-9983783 (WNB),
NASA GSRP grant NGT 5-50247 (AEH),
and the Pennsylvania Space Grant Consortium (AEH).


\begin{references}

\reference{abazajian01}
Abazajian, K., Fuller, G. M., \& Tucker, W. 2001, \apj, 562, 593 

\reference{baldi01}
Baldi, A., Molendi, S., Comastri, A., Fiore, F., Matt, G. \&
Vignali, C.\ 2001, \apj, in press (astro-ph/0108514)

\reference{barger01}
Barger, A. J., Cowie, L. L., Mushotzky, R. F., \&
Richards, E. A.\ 2001, \aj, 121, 662

\reference{brandt01}
Brandt, W.N., et al.\ 2001, \aj, 122, 2810 (B01)

\reference{campana01}
Campana, S., Moretti, A., Lazatti, D., \& Tagliaferri, G.\ 
2001, \apjl, 560, L65 (C01)

\reference{chen97}
Chen, L.-W., Fabian, A. C., \& Gendreau, K. C.\ 1997, \mnras, 285, 449

\reference{della00}
della Ceca, R., Braito, V., Cagnoni, I., \& Maccacaro, T.\
2001 Mem. SAIt in press, (astro-ph/0007430)

\reference{fabian92}
Fabian, A. C. \& Barcons, X.\ 1992, ARA\&A, 30, 429

\reference{field72}
Field, G. B.\ 1972, ARA\&A, 10, 227

\reference{gendreau95}
Gendreau, K.C., et al.\ 1995, \pasj, 47, L5

\reference{giacconi62}
Giacconi, R., Gursky, H., Paolini, F. R., \& Rossi, B. B.\ 1962, 
Phys. Rev. Lett., 9, 439

\reference{giacconi00}
Giacconi, R., et al.\ 2001a, \apj, 551, 624

\reference{giacconi01b}
Giacconi, R., et al.\ 2002, \apjs, in press (astro-ph/0112184) (G02)

\reference{giommi00}
Giommi, P., Perri, M., \& Fiore, F.\ 2000, A\&A, 362, 799

\reference{hasinger98}
Hasinger, G., Burg, R., Giacconi, R., Hartner, G., Schmidt, M.,
Trumper, J., \& Zamorani, G.\ 1998, A\&A, 329, 482

\reference{hasinger01}
Hasinger, G., et al.\ 2001, A\&A, 365, L45

\reference{jauncey75}
Jauncey, D. L.\ 1975, ARA\&A, 13, 23

\reference{marshall80}
Marshall, F., et al.\ 1980, \aj, 235, 4

\reference{murdoch73}
Murdoch, H. S., Crawford, D. F., \& Jauncey, D.\ 1973,
\apj, 183, 1

\reference{mushotzky00}
Mushotzky, R. F., Cowie, L. L., Barger, A. J., \& Arnaud, K. A.\ 2000,
\nat, 404, 459

\reference{schmitt86}
Schmitt, J. H. M. M. \& Maccacaro, T.\ 1986, \apj, 310, 334 

\reference{snow01}
Snowden, S., et al.\ 2001,
http://xmm.gsfc.nasa.gov/docs/xmm/estec2001/ccf

\reference{tozzi01}
Tozzi, P., et al.\ 2001, \apj, 562, 42

\reference{ueda99}
Ueda, Y., et al.\ 1999, \apj, 518, 656

\reference{vecchi99}
Vecchi, A., Molendi, S., Guainazzi, M., Fiore, F., \& Parmar, A.N.\ 1999,
A\&A, 349, L73

\reference{vikhlinin95}
Vikhlinin, A., Forman, W., Jones, C., \& Murray, S., \ 1995,
\apj, 451, 542

\reference{wright94}
Wright, E. L. et al.\ 1994, ApJ, 420, 450

\end{references}
\end{document}